\pdfoutput=1
\documentclass{webofc}
\usepackage[varg]{txfonts}   
\begin{document}
\title{Energy Dependence of Strangeness Production in Heavy-ion Collision}
%
%

\author{\firstname{Md} \lastname{Nasim}\inst{1}\fnsep\thanks{\email{nasim@iiserbpr.ac.in}} }

\institute{
Department of Physics, \\
Indian Institute of Science Education and Research (IISER), Berhampur,\\ Postal Code-760010, Odisha, India.}

\abstract{ An experimental overview of the energy dependence of strangeness production is presented. The strange hadrons are considered a good probe to study the QCD matter created in relativistic nucleus-nucleus collisions.
The heavy-ion experiments at SPS, RHIC, and LHC have recorded a wealth of data in proton-proton, proton-nucleus and nucleus-nucleus collisions at several beam energies.
In this proceeding, I discuss the  invariant yield and azimuthal anisotropy measurement of strange hadrons in nucleus-nucleus collisions at SPS, RHIC, and LHC.}
\maketitle

\section{Introduction}
\label{intro}
 The relativistic heavy-ion collision provides a unique opportunity to study the properties of QCD matter at various temperatures and densities.  In order to map out the phase diagram of the QCD matter, experimental programs started in the early 1990s at the Brookhaven Alternating Gradient Synchrotron (AGS) and the CERN Super Proton Synchrotron (SPS) followed by Relativistic Heavy Ion Collider (RHIC) at Brookhaven and recently at Large Hadron Collider (LHC) at CERN. These experiments allowed us to study the QCD matter over a large range of baryonic chemical potential (or net-baryon density) by varying center-of-mass energy( $\sqrt{s_{NN}}$) of the two colliding nucleus.\\
Strange hadron production in nucleus-nucleus collisions is believed to be an important tool to study the dynamics of the produced QCD matter. Due to the large production cross-section of strange quarks in the QGP, an enhancement in the production of strange hadrons in QGP compared to non-QGP medium is expected~\cite{rafaleski}. Due to low energy density,  the matter produced in the nucleon-nucleon (p+p) collisions is considered as a baseline for the non-QGP medium. Production of strange hadrons is being measured in nucleus-nucleus (A+A) collisions and compared with that in p+p collisions to look for any deviation from the non-QGP baseline. Any significant deviation would hint at the formation of deconfined matter of quarks and gluons in A+A collisions.\\
Among all strange hadrons, multi-strange hadrons (e.g. $\phi$, $\Omega$)  are expected to have relatively small hadronic interaction cross-sections and therefore they can carry the information directly from the chemical
freeze-out stage with almost no distortion due to late-stage hadronic interaction~\cite{star_white,shor}. The measurement of production yield and azimuthal anisotropy of multi-strange hadrons are considered as a useful probe to study the hadronization and collective properties of the QGP medium.\\

\section{Strange Hadron Production}
\label{sec-1}

\subsection{ Transverse momentum ($p_{T}$) integrated particle ratio}
\label{sec-1_a}
\begin{figure}[h]
\includegraphics[scale=0.35]{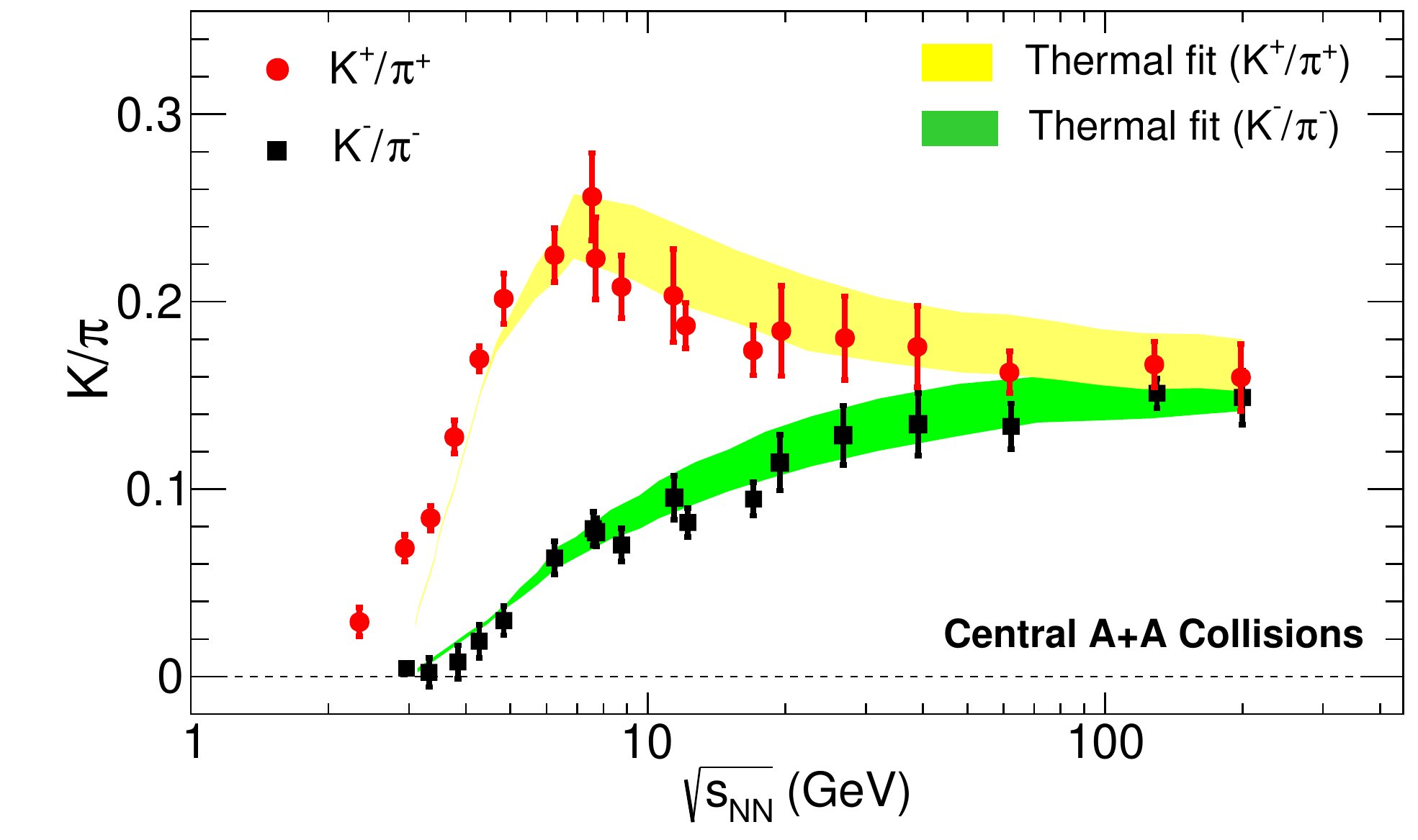}
\includegraphics[scale=0.4]{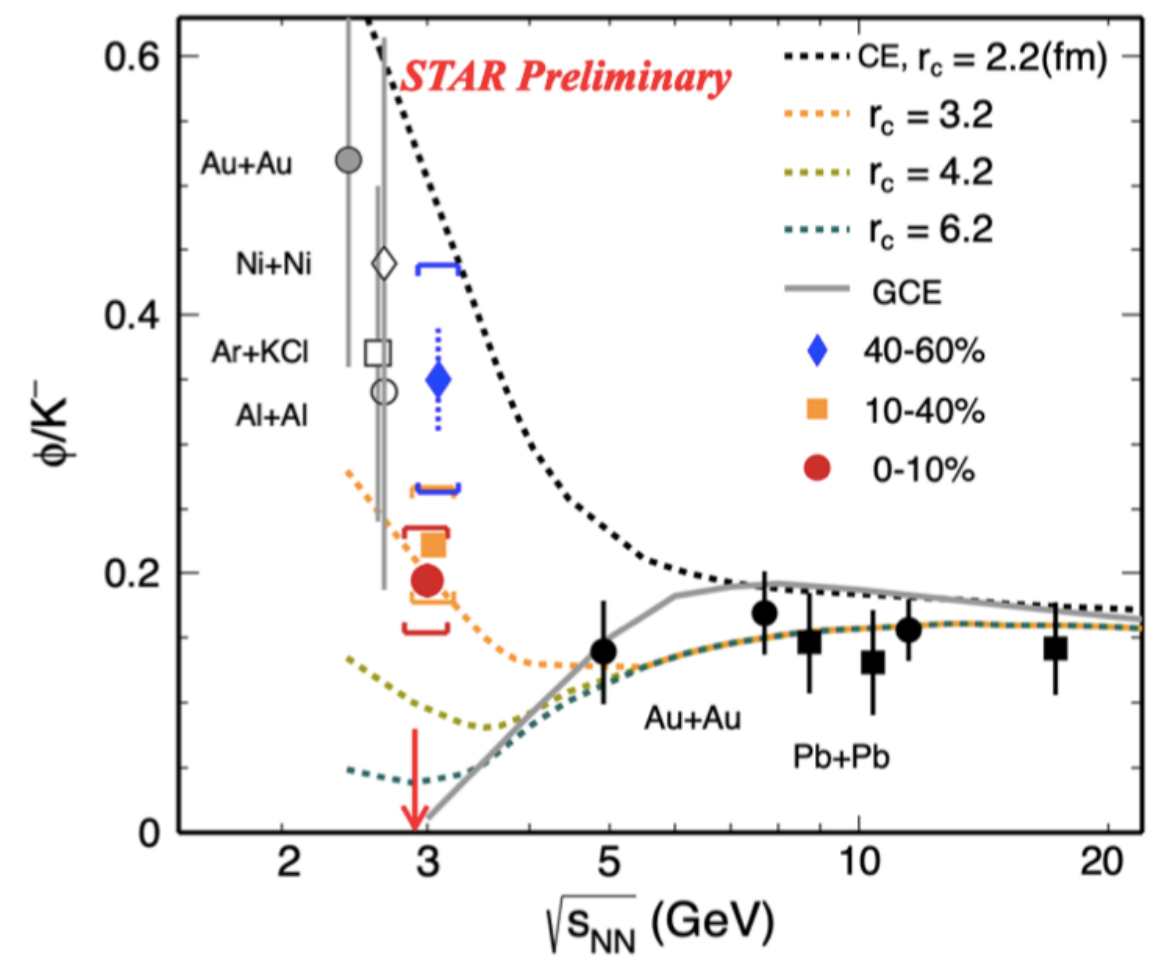}
\caption{Left panel: The $K^{+}/\pi^{+}$ and $K^{-}/\pi^{-}$ ratios as a function of  $\sqrt{s_{NN}}$ in central A+A collision~\cite{AGS_1,AGS_2,AGS_3,SPS_1,SPS_2,RHIC_1}. Right panel: The $\phi/K^{-}$ ratios as a function of  $\sqrt{s_{NN}}$~\cite{HADES,RHIC_3gev}.}
\label{fig-1}       
\end{figure}

\begin{figure}[h]
\centering
\includegraphics[scale=0.3]{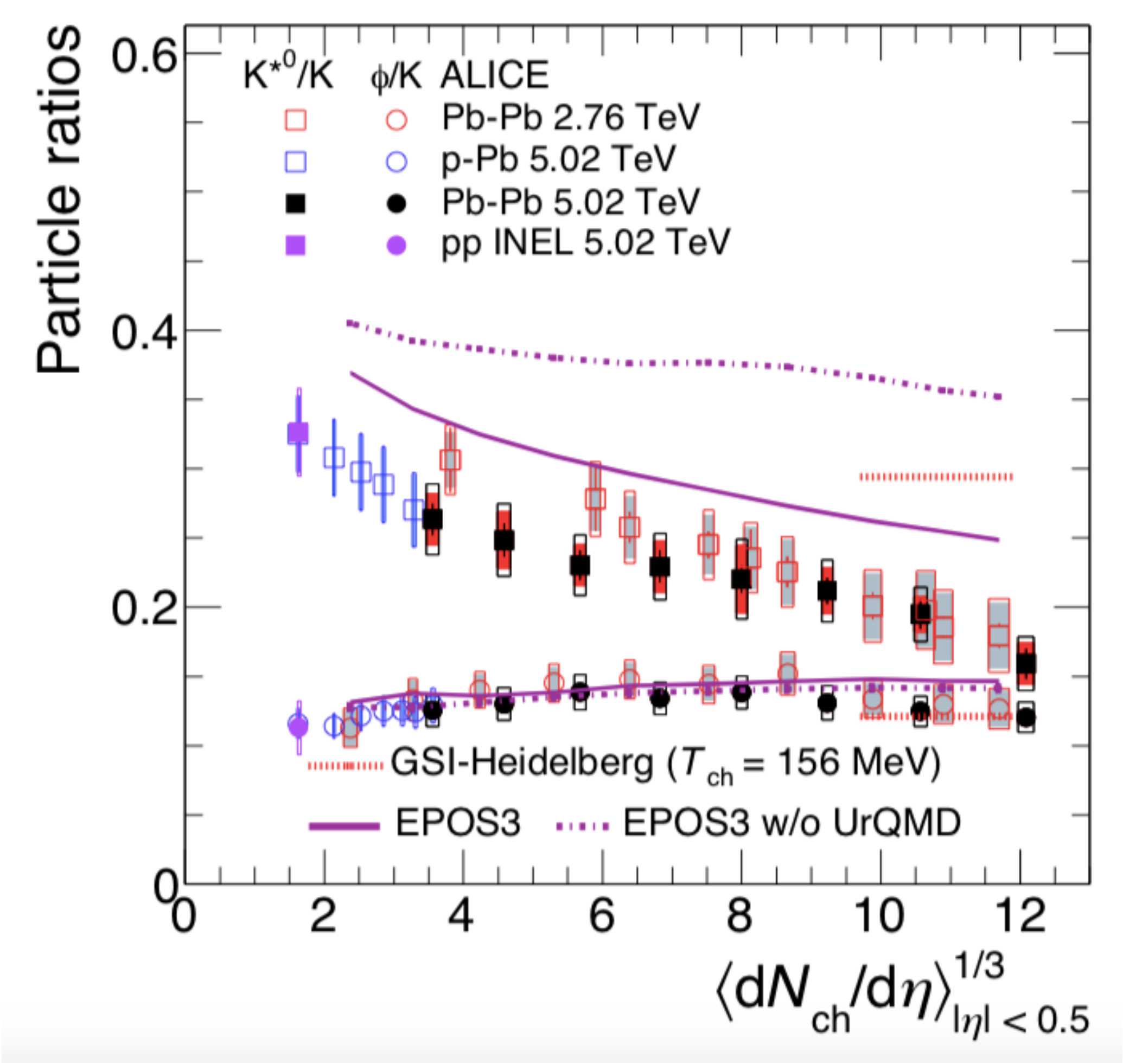}
\caption{The $K^{*0}/K$ and $\phi/K$ ratios as a function of multiplicity in p+Pb and Pb+Pb collisions at 2.76 and 5.02 TeV measured by the ALICE experiment~\cite{LHC_5TeV}.}
\label{fig-2}       
\end{figure}

\textbf{Test of thermal model:}\\
Figure~\ref{fig-1}  (the left panel) shows $K/\pi$ ratios as a function of $\sqrt{s_{NN}}$ in central A+A collision measured  at AGS, SPS, and RHIC~\cite{AGS_1,AGS_2,AGS_3,SPS_1,SPS_2,RHIC_1}. 
The $K/\pi$ ratio is of interest, as it reflects the strangeness content relative to entropy in heavy-ion collisions.  The $K^{+}/\pi^{+}$ and $K^{-}/\pi^{-}$ ratios are shown by red and black symbols, respectively.
The observed peak position in the energy dependence of $K^{+}/\pi^{+}$ has been suggested to be a signature of a change in degrees of freedom~\cite{kpi_horn_1,kpi_horn_2}. 
The thermal model calculations are shown as yellow band for $K^{+}/\pi^{+}$ and green band for $K^{-}/\pi^{-}$~\cite{thermal_1,thermal_2,thermal_3}. It is shown that the thermal model calculation explains the measured $K/\pi$  ratios fairly well.\\
\textbf{Cannonical Ensemble at high-baryon density:}\\
The $\phi/K^{-}$ ratios as a function of $\sqrt{s_{NN}}$ is presented in the right panel of figure~\ref{fig-1}~\cite{HADES,RHIC_3gev}.
The thermal model calculations, based on the Grand Canonical Ensemble (GCE) and Canonical Ensemble (CE) for strangeness with several different choices of strangeness correlation length are shown by different line colors~\cite{thermal_2,thermal_3}.
We can see for $\sqrt{s_{NN}}$ $>$ 5 GeV, both GCE and CE calculations explain the data, however for 
$\sqrt{s_{NN}}$ $\leq$ 3.0 GeV GCE calculation under-predicts the measured  $\phi/K^{-}$ ratios. Calculation using Canonical Ensemble (CE) with strangeness correlation length $\sim$ 3 fm can explain the data measured by STAR and HADES experiment.\\
\textbf{Evidence of hadronic rescattering in central A+A collision:}\\
Figure~\ref{fig-2} shows $K^{*0}/K$ and $\phi/K$ ratios as a function of multiplicity in p+Pb and Pb+Pb collisions at 2.76 and 5.02 TeV measured by the ALICE experiment~\cite{LHC_5TeV}. We see that $K^{*0}/K$ ratios  decrease with increasing multiplicity. Whereas the $\phi/K$ ratio is almost independent of multiplicity. This observation can be understood considering the re-scattering of daughter particles in the hadronic phase. The lifetime of  $K^{*0}$ resonance is short ($\sim$ 4 fm/c) and therefore they decay inside the fireball. The daughter particles of $K^{*0}$ undergo re-scattering in the hadronic medium and their momentum may get changed. This would lead to a loss in the reconstruction of the parent resonance. The lifetime of the $\phi$ meson is $\sim$ 42 fm/c. Because of the longer lifetime, the $\phi$ meson will mostly decay outside the fireball and therefore its daughters will not have much time to rescatter in the hadronic phase. We can also see from figure~\ref{fig-2} that thermal model calculations, which do not include hadronic rescattering, explain $\phi/K$ ratios and over-predict $K^{*0}/K$ at the central Pb+Pb collisions. This is a clear piece of evidence of the hadronic rescattering effect on short-lived resonance in central Pb+Pb collisions.

\subsection{Particle ratio as a function of $p_{T}$}
\label{sec-1_b}
\begin{figure}[h]
\centering
\includegraphics[scale=0.3]{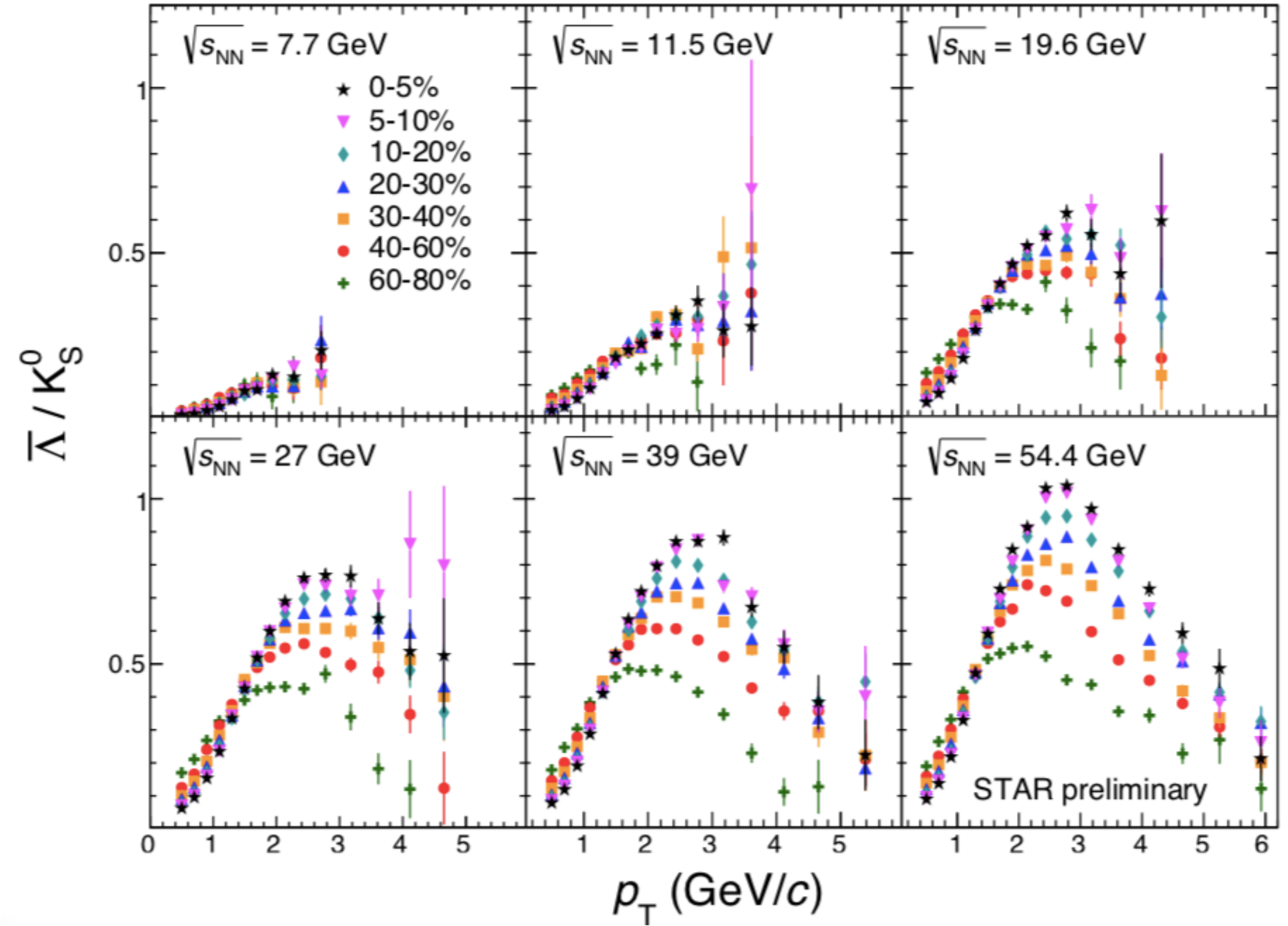}
\caption{The $\bar{\Lambda}/K^{0}_{S}$ ratio as a function of $p_{T}$ for different centralities in Au+Au collisions at $\sqrt{s_{NN}}$ = 7.7–54.4GeV~\cite{RHIC_BES_spectra}.}
\label{fig-3}       
\end{figure}

\textbf{Probing hadronization through baryon-to-meson ratio:}\\
The baryon-to-meson ratios have been predicted to be sensitive to the parton dynamics of the collision system. 
Figure~\ref{fig-3}  shows the $\bar{\Lambda}/K^{0}_{S}$ ratio as a function of $p_{T}$ for different centralities in Au+Au collisions at $\sqrt{s_{NN}}$ = 7.7–54.4 GeV~\cite{RHIC_BES_spectra}.
The enhancement of $\bar{\Lambda}/K^{0}_{S}$  ratios at intermediate $p_{T}$ in central Au+Au collisions compared to peripheral Au+Au at $\sqrt{s_{NN}}$ $\geq$19.6 GeV is interpreted as a consequence of hadron formation through parton recombination.We do not see any significant difference between central and peripheral collisions for $\sqrt{s_{NN}}$ $\leq$ 11.5 GeV. This may indicates formation of hadronic interaction dominated matter  at $\sqrt{s_{NN}}$ $\leq$ 11.5 GeV.

\subsection{Nuclear modification factor}
\label{sec-1_c}
\begin{figure}[h]
\centering
\includegraphics[scale=0.3]{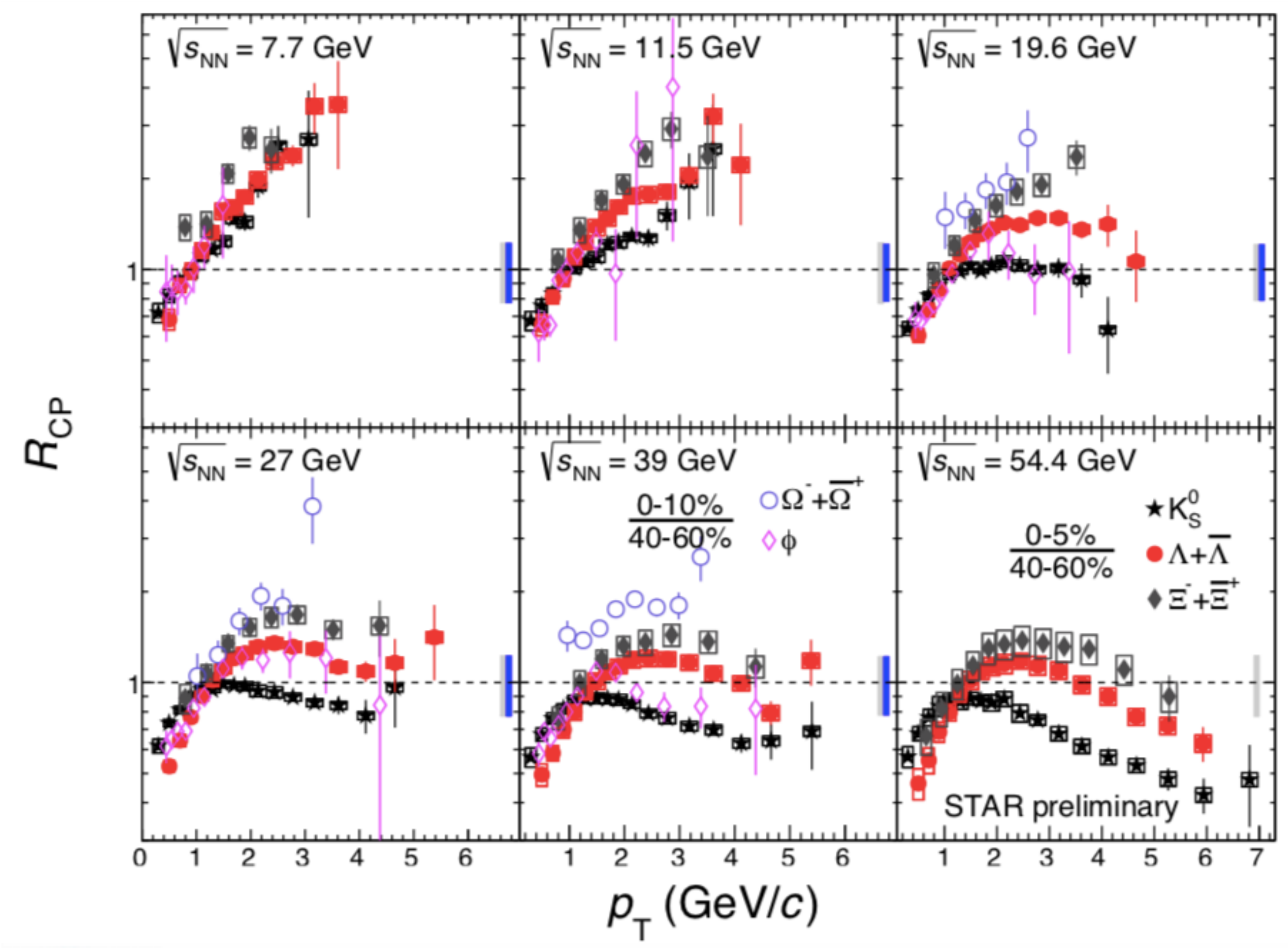}
\caption{The nuclear modification factor of strange hadrons measured in Au+Au collisions at $\sqrt{s_{NN}}$ = 7.7–54.4GeV~\cite{RHIC_BES_spectra}. }
\label{fig-4}       
\end{figure}

\textbf{Probing QGP through parton energy loss:}\\
The nuclear modification factor (R$_{CP}$) is defined as the ratio of the particles yield in the central to peripheral collisions normalized by the number of binary collisions (Nbin). The value of Nbin is calculated from the Monte Carlo Glauber simulation.  If R$_{CP}$ is equal to one, then the nucleus-nucleus collision is simply  the superposition of nucleon-nucleon collisions. Deviation of R$_{CP}$ from the unity would imply contribution from the nuclear medium effects. The measured R$_{CP}$, which is less than one at high $p_{T}$, at top RHIC and LHC energies indicates the energy loss of partons traversing through QGP medium.  Figure~\ref{fig-4} show R$_{CP}$ of strange hadrons measured in Au+Au collisions at $\sqrt{s_{NN}}$ = 7.7–54.4GeV~\cite{RHIC_BES_spectra}. From figure~\ref{fig-4} one can see, at the intermediate $p_{T}$ , R$_{CP}$ goes above unity with the decrease in beam energy. This indicates that at lower beam energy the parton energy loss effect could be less important.

\section{Collective Flow of Strange Hadrons}
\label{sec-2}

\begin{figure}[h]
\centering
\includegraphics[scale=0.55]{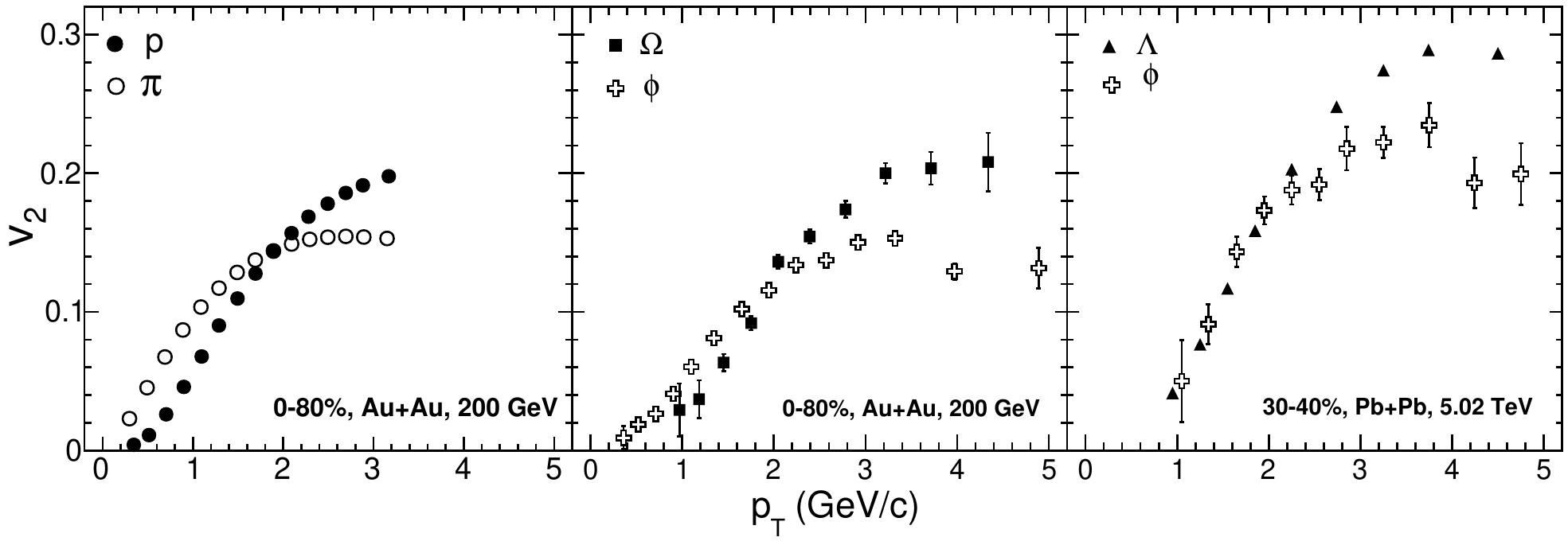}
\caption{The elliptic flow as a function of $p_{T}$ for $\pi$, $p$, $\phi$,
and $\Omega$ for 0\%–80\% centrality in Au+Au collisions at $\sqrt{s_{NN}}$ = 200 GeV and for $\phi$, and $\Lambda$ in 30\%–40\% central Pb+Pb collisions $\sqrt{s_{NN}}$ = 5.02 TeV~\cite{RHIC_200_v2,LHC_5TeV_v2}.}
\label{fig-5}       
\end{figure}

\begin{figure}[h]
\includegraphics[scale=0.3]{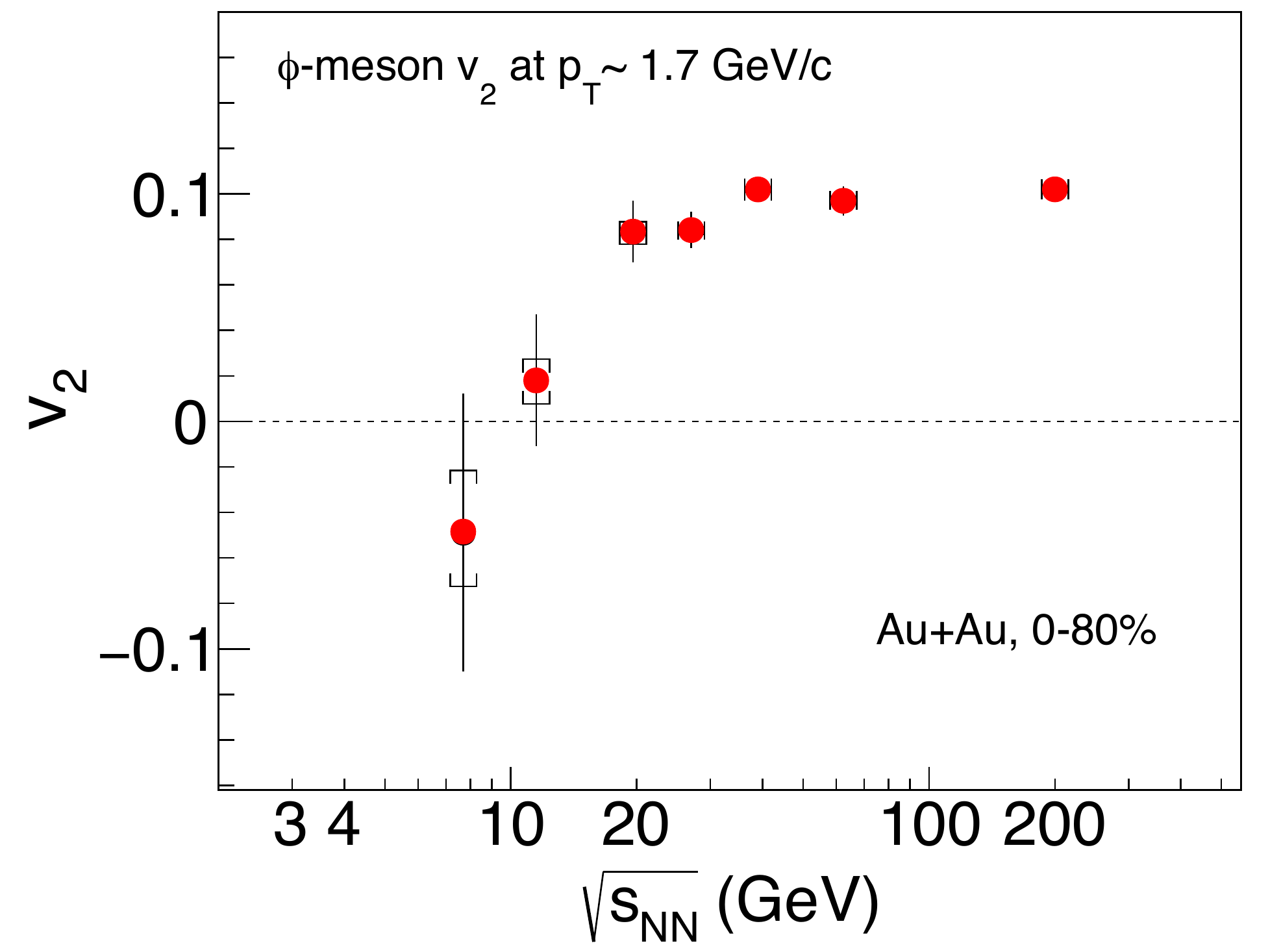}
\includegraphics[scale=0.33]{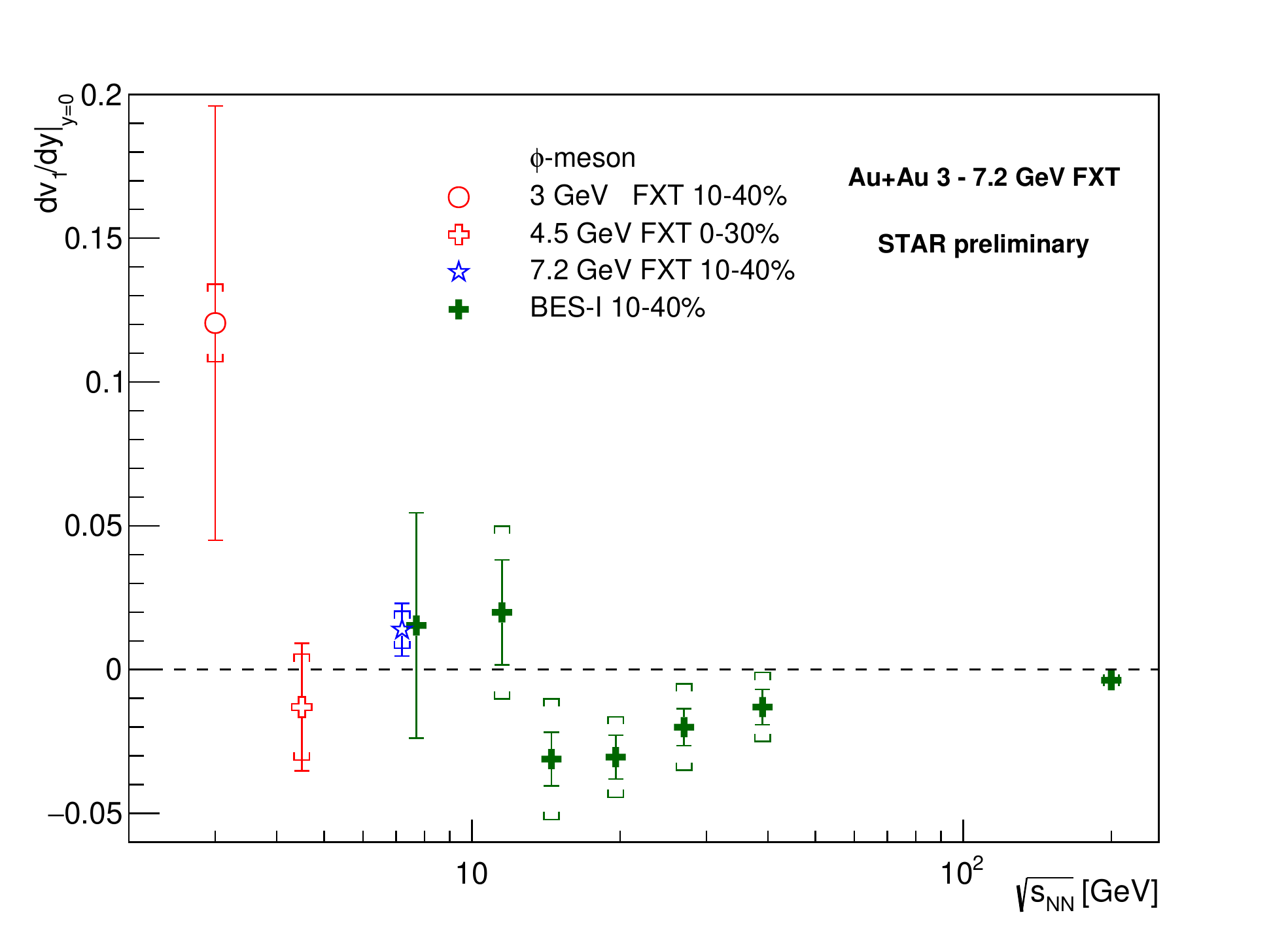}
\caption{The energy dependence of $\phi$ $v_{2}$ at intermediate $p_{T}$ and  slope of directed flow ($dv_{1}/dy$) measured by the STAR experiment~\cite{RHIC_BES_v2,RHIC_BES_v1}.}
\label{fig-6}       
\end{figure}

\textbf{Partonic collectivity at top RHIC and LHC energies:}\\
Figure~\ref{fig-5}  shows the  elliptic flow coefficient ($v_{2}$) as a function of $p_{T}$ for $\pi$, $p$, $\phi$,
and $\Omega$ for 0–80\% centrality in Au+Au collisions at $\sqrt{s_{NN}}$ = 200 GeV and for $\phi$, and $\Lambda$ in 30–40\% central Pb+Pb collisions at $\sqrt{s_{NN}}$ = 5.02 TeV~\cite{RHIC_200_v2,LHC_5TeV_v2}. We see mass ordering in $v_{2}$ at low $p_{T}$ and particle type dependence at the intermediate $p_{T}$. The observed mass ordering can be explained by hydrodynamics models whereas particle type dependence at the intermediate $p_{T}$ can be explained through the parton recombination model. It is clear from figure~\ref{fig-5} that the $v_{2}$ of hadrons consisting of only strange quarks ($\phi$ and $\Omega$) is similar to that of non-strange hadrons $\pi$ and $p$. As we know that the  $\phi$ and $\Omega$ do not participate strongly in the hadronic interactions unlike $\pi$ and $p$, therefore figure~\ref{fig-5} suggests that the major part of collectivity is developed during the partonic phase at top RHIC energy. This is also true for LHC energies as shown by the ALICE experiment~\cite{LHC_5TeV_v2}. \\
\textbf{Turn-off of QGP  and change of equation of states ?}\\
The energy dependence of $\phi$ meson $v_{2}$ measured at intermediate $p_{T}$ is shown in figure~\ref{fig-6}~\cite{RHIC_BES_v2}. We find that the magnitude of $v_{2}$ of $\phi$ is almost constant ($\sim$0.1) for  $\sqrt{s_{NN}}$ $\geq$ 19.6 GeV. However, for $\sqrt{s_{NN}}$ =11.5 and 7.7 GeV, magnitude  of $\phi$ $v_{2}$ is small or close zero within large uncertainties. Small  $\phi$ $v_{2}$ is considered as a signature of formation of hadronic interaction dominated matter as $\phi$ mesons are expected to have small hadronic interaction cross sections.
The energy dependence of  the slope of $\phi$  directed flow ($v_{1}$) is shown in the right panel of figure~\ref{fig-6}~\cite{RHIC_BES_v1}. There are indications of change in the sign of slope of  $v_{1}$ of $\phi$ below $\sqrt{s_{NN}}$ = 10 GeV. This could be related to the change of equation of state at the low beam energy.
The uncertainties in the current measurement are large, upcoming BES-II program at RHIC is expected to provide precision measurement of $\phi$  $v_{1}$  and $v_{2}$. 

\section{Summary}
\label{summa}
The energy dependence of strange hadron production in heavy-ion collisions is presented. It is found that  the thermal model explains the measured particle ratios (e.g. $K/\pi$ and $\phi/K$) and its energy dependence.
The ratio $\phi/K$ is found to be a sensitive probe to study the strangeness canonical suppression.  We observed thermal model calculations over-predicts $K^{*0}/K$ at the central Pb+Pb collisions. Also, the $K^{*0}/K$  ratio decreases with increasing event multiplicity. This clearly indicates the effect of hadronic re-scattering on short-lived resonance in central  A+A collisions. 
The measured elliptic flow of strange hadrons found to be similar to that of non-strange hadrons at top RHIC and LHC energies, indicating formation of partonic collectivity.\\
The measure $\phi$ $v_{2}$, at intermediate $p_{T}$, is found to be close zero within large uncertainties at $\sqrt{s_{NN}}$ =11.5 and 7.7 GeV. The $\phi$  $v_{1}$ slope changes sign below $\sqrt{s_{NN}}$ = 10 GeV. The nuclear modification factor is greater that unity at the intermediate $p_{T}$ for 
$\sqrt{s_{NN}}$ $\leq$ 19.6 GeV. The baryon-to-meson  ($\bar{\Lambda}/K^{0}_{S}$) ratios as a function of $p_{T}$ is found to be centrality independent for $\sqrt{s_{NN}}$ =11.5 and 7.7 GeV. 
These results indicate the formation of hadronic interaction dominated matter below $\sqrt{s_{NN}}$ = 20 GeV. 
%

\begin{thebibliography}{}
%
%

\bibitem{rafaleski}  J. Rafelski and B. Muller, Phys. Rev. Lett.  48, 1066 (1982).

\bibitem{star_white} J. Adam  et al. (STAR),  Nucl. Phys. A 757, 102 (2005).

\bibitem{shor} A. Shor, Phys. Rev. Lett. 54, 1122 (1985).

\bibitem{AGS_1} L. Ahle et al. (E866, E917), Phys. Lett. B 476, 1 (2000).
\bibitem{AGS_2} L. Ahle et al. (E-802, E-866), Phys. Rev. C 60, 044904 (1999).
\bibitem{AGS_3}  L. Ahle et al. (E866, E917), Phys. Lett. B 490, 53 (2000).
\bibitem{SPS_1} S. Afanasiev et al. (NA49), Phys. Rev. C 66, 054902 (2002). 
\bibitem{SPS_2} C. Alt et al. (NA49), Phys. Rev. C 77, 024903 (2008).
\bibitem{RHIC_1} L. Adamczyk et al. (STAR), Phys. Rev. C 96, 044904 (2017) and sQM2021 proceedings.


\bibitem{kpi_horn_1} M. Gazdzicki and M. I. Gorenstein, Acta Phys. Polon. B 30, 2705 (1999). 
\bibitem{kpi_horn_2} J. Cleymans, et al. Phys. Lett. B 615, 50 (2005).

\bibitem{thermal_1} J. Randrup and J. Cleymans, Phys. Rev. C 74, 047901 (2006).
\bibitem{thermal_2} A. Andronic, et al. Nucl. Phys. A 772, 167 (2006)
\bibitem{thermal_3} J. Cleymans et al. Phys.Lett. B 603, 146 (2004).

\bibitem{HADES}  G. Agakishiev et al. (HADES) Phys. Rev. C 80, 025209 (2009)
\bibitem{RHIC_3gev} M. S. Abdallah et al. (STAR), arXiv:2108.00924 and sQM2021 proceedings.

\bibitem{LHC_5TeV}  S. Acharya, et al. ( ALICE) PLB 802, 135225 (2020).

\bibitem{RHIC_BES_spectra} J. Adam  et al. (STAR), Phys. Rev. C 102, 34909 (2020) and sQM2021 proceedings.
 
\bibitem{RHIC_200_v2}  L. Adamczyk et al. (STAR),  Phys. Rev. Lett. 116,  062301 (2016).

\bibitem{LHC_5TeV_v2}    S. Acharya, et al. ( ALICE)  JHEP 09, 006 (2018).

\bibitem{RHIC_BES_v2} L. Adamczyk et al. (STAR), Phys. Rev. Lett. 110, 142301 (2013)

\bibitem{RHIC_BES_v1} L. Adamczyk et al. (STAR), Phys. Rev. Lett. 120, 62301 (2018) and sQM2021 proceedings.
 
 
\end{thebibliography}
%
%

\end{document}